**Accurate determination of band tail properties in amorphous semiconductor thin film with Kelvin Probe Force Microscopy**

Luca Fabbri, Camilla Bordoni, Pedro Barquinha, Jerome Crocco, Beatrice Fraboni, Tobias Cramer

Luca Fabbri, Camilla Bordoni, Prof. Beatrice Fraboni, Prof. Tobias Cramer

Department of Physics and Astronomy

University of Bologna

Viale Berti Pichat 6/2, 40127, Bologna, Italy

Prof. Pedro Barquinha

CENIMAT/I3N, Departamento de Ciência dos Materiais (DCM), and CEMOP/UNINOVA

NOVA School of Science and Technology (FCT NOVA), Universidade NOVA de Lisboa,

2829-516 Caparica, Portugal

Jerome Crocco

DPIX, Chief Technology Officer

1635 Aeroplaza Dr, Colorado Springs, 80916, US

Email: tobias.cramer@unibo.it

**Abstract**

Amorphous oxide semiconductors are receiving significant attention due to their relevance for large area electronics. Their disordered microscopic structure causes the formation of band tails in the density of states (DOS) that strongly affect charge transport properties. Bandtail properties are crucial to understand for optimizing thin film device performance. Among the available techniques to measure the DOS, KPFM is exceptional as it enables precise local electronic investigations combined with microscopic imaging. However, a model to interpret KPFM spectroscopy data on amorphous semiconductors of finite thickness is lacking. To address this issue, we provide an analytical solution to the Poisson's equation for a metal-insulator-semiconductor (MIS) junction interacting with the AFM tip. The solution enables us to fit experimental data for semiconductors with finite thickness and obtain the DOS parameters, such as band tail width ($E_t$), doping density ($N_D$), and flat band potential. To demonstrate our method, we perform KPFM experiments on Indium-Gallium-Zinc Oxide (IGZO) thin film transistors (IGZO-TFTs). DOS parameters compare well to values obtained with photocurrent spectroscopy. We demonstrate the potentials of the developed method by investigating the impact of ionizing radiation on DOS parameters and TFT performance. Our results provide clear evidence that the observed shift in threshold voltage is caused by static charge in the gate dielectric leading to a shift in flat band potential. Band-tails and doping density are not affected by the radiation. The developed methodology can be easily translated to different semiconductor materials and paves the way towards quantitative microscopic mapping of local DOS parameters in thin film devices.

**Introduction**

Amorphous semiconductors are fundamental for large area electronic devices such as active matrices in detectors or displays. The disordered structure of the amorphous phase causes the formation of localized states at the edges of the valence and conduction band, called band tails. Such band tails are of fundamental importance for the transport properties of the material since they act as trapping states for carriers.[1] An important technological aim is the downscaling of amorphous semiconductor device dimensions where local border effects become increasingly relevant. In such a situation band-tails and general DOS properties have to be understood as a local property which is influenced by local variations in microsctructure, chemical composition or electrostatic charge trapping. However, most techniques that assess DOS properties rely on macroscopic probes and quantify average material properties. Examples are spectroscopic methods that measure absorption or photocurrent generation as well as electrical methods based capacitance or transistor measurements and modelling.[2–7] Instead, improved device understanding and optimization requests for a microscopic technique that could reveal DOS properties with nanometric resolution on working microstructured semiconductor thin film devices.

Kelvin Probe Force Microscopy (KPFM) is an advanced atomic force microscopy technique that can measure electrical properties of a sample with high resolution and precision, including the density of states.[8–11] With KPFM one measures the contact potential difference between the tip and the sample[12], but band bending and non-uniform charge distribution can complicate the interpretation of results.[13] However, KPFM has been successfully applied to measure surface potential in thin film transistors[14] and to determine the DOS in ultrathin semiconducting layers.[15,16] For instance, Roelofs et al demonstrated quantitative measurements in a molecular semiconducting layer build by selfassembly.[15] Zhang et al demonstrated the formation of trap states at the $SiO_2-$ Oligothiophene monolayer interface and obtained the doping density and sign of charge carriers using KPFM.[16] For semiconducting layers of finite thickness, the interpretation of KPFM measurements is complicated by band bending and the Poisson-equation has to be taken into account.[17] In their study of disordered oxide semiconductors, Germs et al. proposed an iterative numerical procedure to approximate the DOS and could demonstrate an exponential shaped DOS for these materials.

[18] Despite this progress, a reliable model to derive the DOS from KPFM measures in amorphous semiconductors with finite layer thickness is not available.

One of the most promising amorphous semiconductor is Indium Gallium Zinc Oxide (IGZO) as it can be easily deposited by sputtering and presents a mobility of the order of $10\ cm^2/Vs$ which is ten times higher than amorphous silicon.[19] IGZO presents a reduced band tail width with respect to a-Si and so an increased mobility.[20] Band tails in IGZO can be modeled as exponentials described by two parameters: band tail width and oxygen vacancies doping density. This assumption is confirmed by experimental studies.[1,2,21] In addition to the high mobilities, IGZO is characterized by larger bandgap than a-Si causing a low off-state and leakage current and high transparency. These properties make IGZO it an ideal material for use in transparent electronics such as flexible displays and sensors.[19] Investigations of the DOS properties of IGZO have been performed by several groups, employing a wide range of techniques.[2,3,21]

Here we demonstrate the determination of IGZO band tail parameters with KPFM spectroscopic measurements on thin film transistors. Our approach is based on the solution of Poisson's equation for a system that comprises a metal-insulator-semiconductor (MIS) junction interacting with an AFM tip. The mathematical model considers both band bending and charge accumulation at the interface between the semiconductor and the insulator. We provide both the full numerical solution and an approximated analytical one. The model provides access to the band-tail width, the doping density and the flatband potential of the semiconductor. We compare the results from KPFM measurements with photocurrent spectroscopy and transistor measurements. We further demonstrate the value of our method by analyzing the effect of ionizing radiation on DOS properties.

**Results**

The standard configuration for KPFM experiments on thin film semiconductor devices is depicted in Figure 1a. The KPFM tip is operated above the exposed semiconducting thin film. Source and the drain electrodes contact the semiconducting layer and field effect is exerted by the gate electrode separated from the semiconductor by a thin dielectric layer. The source is connected to ground while potentials $V_D$ and $V_G$ applied to the drain and the gate are controlled by a source measure unit. The KPFM experiment probes the local surface morphology of the semiconducting channel and measures the local surface voltage $V_{KPFM}$. Depending on the applied gate voltage, two limiting cases of the surface voltage have to be considered: In depletion the channel is void of carriers and behaves as a dielectric, the electric field of the gate penetrates unscreened the semiconductor and interacts with the KPFM tip. The surface voltage varies with the applied gate voltage. In contrast, in accumulation the semiconducting layer behaves as a conductor and its carriers screen the gate field. Accordingly, surface voltage is constant. KPFM spectroscopy experiments that aim at probing the DOS of the semiconductor characterize the surface voltage as a function of the gate voltage while the transition from depletion to accumulation occurs.

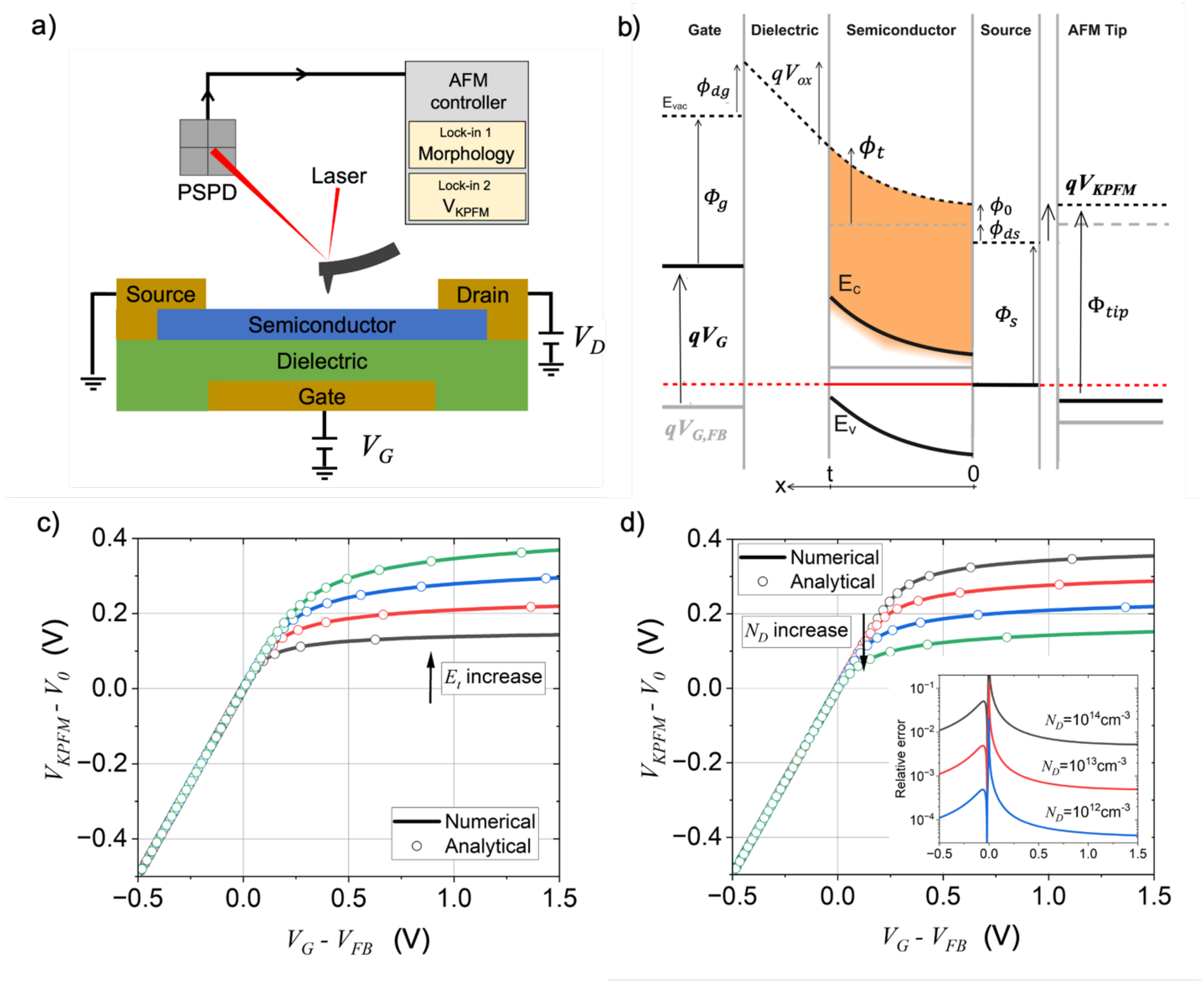

**Figure 1: KPFM spectroscopy and model (a) Schematic of the Kelvin-Probe-Force Microscopy experiment performed on working thin film transistors. In the KPFM spectroscopy experiments, the tip potential is measured while the transistor gate voltage is varied. (b) Schematic band diagram of the metal – insulator – semiconductor junction interacting with the AFM tip. Grey color corresponds to the flat band condition. Energy levels in depletion are depicted in black. The semiconductor is grounded through the source. The AFM tip is interacting with the semiconductor. (c) and (d) show simulated KPFM spectroscopy measurements. Solution to Poisson's equation are shown for different values of band tail width $E_t$ ranging from 20meV to 50meV with fixed doping densities $N_D = 10^{12} cm^{-3}$ in (c) and different values of $N_D$ ranging from $10^{11}$cm$^{-3}$ to $10^{14}$cm$^{-3}$ and fixed $E_t =$**

**$30meV$ in (d). The inset shows the calculated error between numerical and analytical solution for different traps densities.**

In order to derive a quantitative model for the KPFM experiments on the semiconducting thin film, we consider the schematic energy diagram of the MIS junction during the interaction with the AFM tip as shown in Figure 1b. All relevant energetic parameters are indicated in the diagram. We distinguish two cases: first, the flat band condition that is drawn in grey lines. Second, the depletion condition as highlighted with black lines. Solid lines refer to energy levels, while dashed lines stand for the vacuum level. The Fermi Level in the semiconductor is given by the grounded source electrode and is colored in red in the diagram. The application of a voltage to the gate produces a band bending inside the semiconductor. This effect is quantified with the potential $\phi(x)$ which expresses the shift of vacuum level in the semiconductor with respect to flat band condition. The spatial coordinate of $\phi(x)$ starts at the external interface $\phi(x=0)=\phi_0$ and ends at the interface with the gate dielectric $\phi(x=t)=\phi_t$. Here $t$ denotes the thickness of the semiconducting layer. Other important parameters in the diagram regard the drop in vacuum level across the gate dielectric $V_{ox}$ and the work functions of the gate and source electrodes, $\Phi_g$ and $\Phi_s$, respectively. Additional offsets in vacuum level occur at interfaces due to static trapped charges or interfacial dipoles. They are denoted as $\phi_{ds}$ and $\phi_{dg}$ and refer to the interfaces source/semiconductor and gate/dielectric, respectively. Finally, the diagram includes the AFM probe interacting electrostatically with the MIS junction. The potential $V_{KPFM}$ is controlled by the KPFM feedback loop targeting the absence of electrostatic forces between tip and sample. Accordingly, there is no electric field between the tip and the semiconductor surface, and the vacuum level of the AFM tip aligns to the semiconductor interface. The measured $V_{KPFM}$ thus reports on changes in vacuum level at the external semiconductor interface occurring when the gate voltage is changed.[13] Based on the energy diagram we derive the quantitative model for the KPFM experiment in two steps. First, we develop the relation for the potential $\phi(x)$ describing band bending in an amorphous semiconductor of finite thickness. Second, we identify how the experimental observables $V_{KPFM}$ and $V_G$ are linked to the parameters entering the potential function $\phi(x)$.

Band bending occurs due to the buildup of local space charge $\rho(x)$ in the semiconductor. For a n-doped amorphous semiconductor we can write

$$\rho(x) = q_0 N_D \left[1 - \exp\left(\frac{\phi(x)}{E_t}\right)\right] \tag{1}$$

where $E_t$ describes the exponential band tail[20,22], $DOS(E) \propto exp(E/E_t)$, $N_D$ introduces the concentration of n-type doping sites and $q_0$ refers to the unit charge. Equation (1) is derived from the solution of the Fermi-Dirac integral under the zero temperature assumption. Finite temperatures can be accounted for as a temperature dependent shift in flat band voltage. The equation for $\phi(x)$ is then defined by combining (1) with the Poisson equation

$$\frac{d^2\phi(x)}{dx^2} = -\frac{\rho(x)}{\varepsilon\varepsilon_0} \tag{2}$$

The resulting differential equation requires two boundary conditions to provide a unique solution for $\phi(x)$. Following our KPFM experiment, we consider the external semiconductor interface at x=0 where we nullify the electric field and measure the local surface potential. Accordingly, we set $\phi(x=0) = \phi_0$ and $\phi^{\prime}(x=0) = 0$. For these boundary conditions we obtain an analytical solution for the condition $\phi(x=0) \gg E_t$:

$$\phi(x) = E_t \log\left\{exp\left(\frac{\phi_0}{E_t}\right)\left(tan^2\left[\frac{x}{\lambda\sqrt{2}}exp\left(\frac{\phi_0}{2E_t}\right)\right] + 1\right)\right\} \tag{3}$$

Here $\lambda = \sqrt{E_t\varepsilon\varepsilon_0/q_0^2 N_D}$ describes a characteristic length scale of the semiconductor. The solution describes the band bending occurring when negative carriers accumulate in the channel and strongly screen the gate field and depends explicitly on the DOS parameters. Below we compare the analytical solution with the complete numerical one and show that it is in many situations also a good approximation to describe KPFM experiments in depletion regime.

In the second part of our derivation, we link the potential function $\phi(x)$ to the experimental observables $V_G$ and $V_{KPFM}$. Key to this are the function values at the external and internal semiconductor interfaces $\phi(x=0) = \phi_0$ and $\phi(x=t) = \phi_t$, respectively. When a potential is applied to the gate, the vacuum level and the conduction band edge are shifted and, due to finite thickness of the semiconductor, a residual of the gate electric field can reach the external semiconductor interface. The resulting shift in the conducting band is $\phi_0$ and it is directly related to the voltage $V_{KPFM}$ measured at the AFM tip in the KPFM experiment. A direct measurement of $\phi_0$ is difficult as different offsets contribute to $V_{KPFM}$ such as the work function of the AFM-tip, possible dipole layer forming at the semiconductor interface and amplifier offsets in the KPFM signal acquisition. Accordingly, we introduce the kelvin probe surface voltage measured in flat band conditions $V_{KPFM,\ FB}$ and obtain

$$\phi_0 = V_{KPFM} - V_{KPFM,FB} \tag{4}$$

Next we consider the semiconductor interface with the gate dielectric and derive the relation that links the surface voltage $\phi_t$ to the gate voltage $V_G$. Following the energy diagram in figure 1b we find

$$qV_G = qV_{G,FB} + \phi_t + qV_{ox} \tag{5}$$

Here $V_G$ is the flat band voltage and $V_{ox}$ is the potential drop over the gate dielectric, which is given by $V_{ox} = \sigma/c_{ox}$. The capacitance of the gate dielectric $c_{ox}$ is a constant and the surface charge density in the semiconducting channel $\sigma$ can be directly computed from the potential function $\phi(x)$ (see Suppl. Mat.). We note that equation (3) allows to express both $\phi_t$ as well as $\sigma$ as a function of $\phi_0$. Accordingly, the combination of equations (3), (4) and (5) provides an analytical expression for $V_G$ as a function of $V_{KPFM}$.

Figures 1c and 1d show simulated $V_{KPFM}$ vs $V_G$ curves for different combinations of DOS parameters. At negative gate voltages, when the channel is in depletion, a linear relation with unitary slope is observed as the gate field passes through the semiconductor. Instead towards positive gate voltages the curve flattens and reaches a plateau. $V_{KPFM}$ becomes independent on $V_G$ as the gate field is fully screened in the semiconducting channel. Figure 1c demonstrates how the parameter $E_t$, describing the width of the conduction band tail, strongly impacts on the shape of the curves. Wider band tails (larger $E_t$) cause a wider transition from full depletion to full accumulation and hence a reduced curvature of the $V_{KPFM}$ vs $V_G$ curve. Accordingly, the plateau is reached at higher KPFM voltages. In contrast, the parameter describing the dopant concentration $N_D$ has a different effect as shown in Figure 1d. The shape of the $V_{KPFM}$ vs $V_G$ curve is not altered with $N_D$, but higher values cause a shift of the curve to lower values of $V_{KPFM}$. Due to the high dopant concentration, more charge carriers are in the channel at flat band potential and full screening of the gate field occurs already at smaller shifts in vacuum level. The two graphs also show the excellent agreement between the numerical and approximated analytical solutions for the tested combinations of DOS parameters. Interestingly for a wide range of DOS parameters, the analytical solution turns out to be also a good approximation in depletion condition even though the condition $\phi \gg E_t$ is no longer valid. The relative error of the approximate analytical solution is largest in flat band condition and increases at higher doping concentrations $N_D$.

The simulations show that the band tail width $E_t$ can be determined from experimental data as it impacts directly on the shape of the $V_{KPFM}$ vs $V_G$ curves. In contrast, the doping density $N_D$ is not uniquely determined by the curve. Changes in $N_D$ only translate the curve and variations in the flat band offset parameters $V_{G,FB}$ and $V_{KPFM,FB}$ have the same effect. In order to determine $N_D$ additional data needs to be considered and here we describe two possible approaches: first, flat band voltage $V_{G,FB}$ can be measured with alternative techniques such as capacitance measurements. Subsequently also $V_{KPFM,FB}$ and $N_D$ can be obtained from the data. However, capacitance measurements are not always feasible and border effects in small transistor channels can make their interpretation difficult. Therefore, we follow a second approach that aims at determining $V_{KPFM,FB}$ with additional KPFM measurements done on the source electrode. Following the energy diagram in figure 1b, one can write $V_{KPFM,FB} = V_{KPFM,s} + \phi_{ds}$. Here $\phi_{ds}$ is the barrier that charge carriers have to overcome when injection or extraction occurs from the semiconducting channel. As transport in optimized TFTs is not limited by contact effects as shown experimentally, we can assume that $\phi_{ds} < k_B T$ and exploit this assumption in the determination of $V_{KPFM,FB}$ and subsequently $N_D$. The measurement of $V_{KPFM,s}$ is easily achieved in the same experiment as it is enough to enlarge the scanning area to include part of the source electrodes.

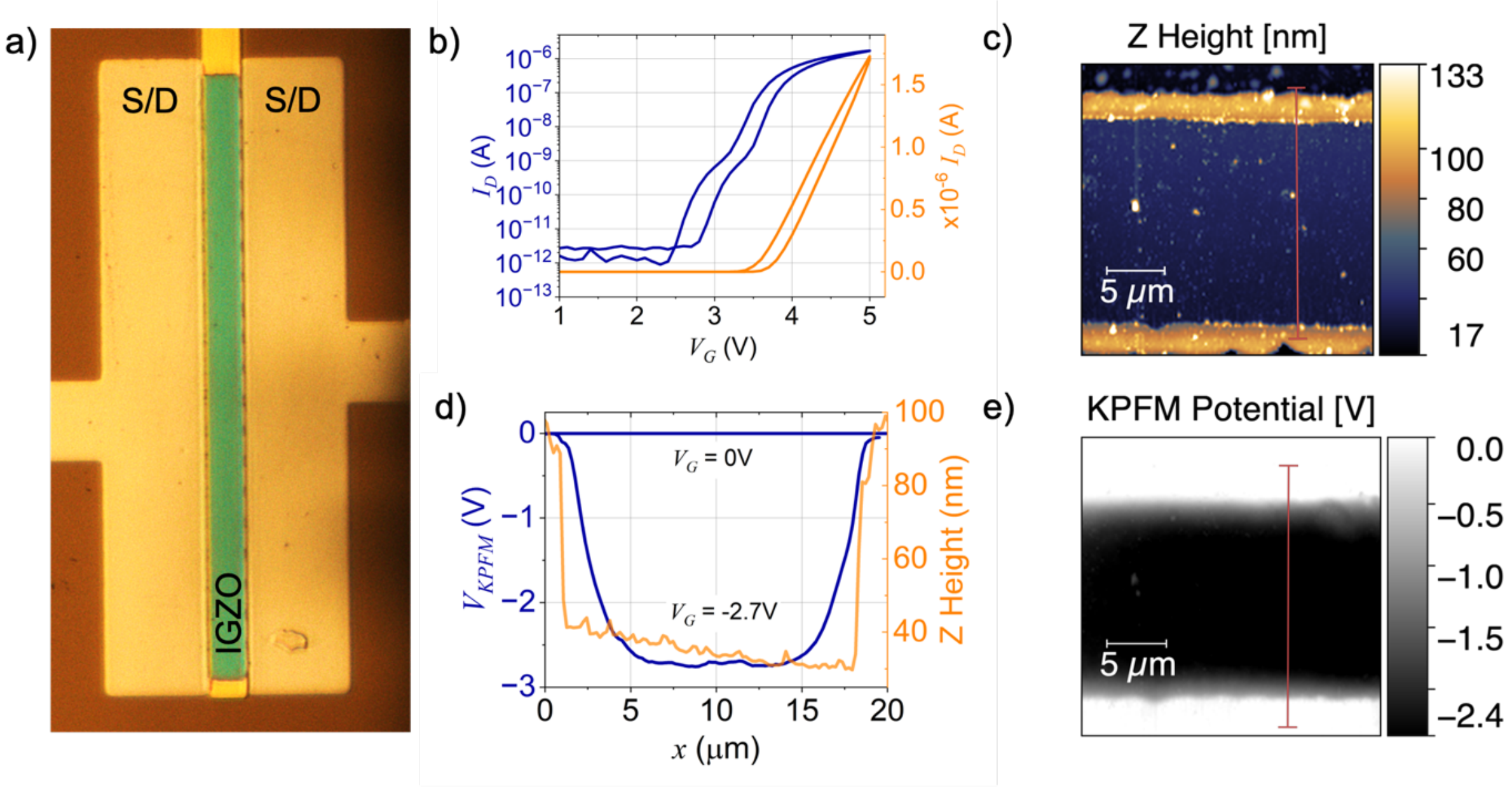

**Figure 2: (a) Optical microscope image of a a-IGZO thin film transistor with channel's geometry 320x20 μm. Source and drain electrodes are deposited on the top of the device while the gate is under the semiconducting channel. (b) Transfer characteristic in log scale (blue) and linear scale (orange). (c) AFM height map of the channel. (d) KPFM potential and morphology profiles of the sample. The profile line is identified with the red marker in subfigures (c) and (e). (e) KPFM potential map of of the channel at $V_G = -2.7$V.**

To test our model, we perform KPFM experiments on indium gallium zinc oxide (IGZO) thin film transistors with a 60nm thick semiconducting layer and a dielectric composed of 80nm of $Ta_2O_5$ and 20nm of $Al_2O_3$. The source and drain contacts are made of molybdenum as well as the gate electrode. The channel has a geometry of $W=320\mu m$ and $L=20\mu m$ The optical microscope image of the device is reported in figure 2a. In this device the gate is placed below the channel while the source and drain electrodes are deposited on the top, as reported in figure 2a. Figure 1b shows the transfer characteristic of the device. We measured a mobility of $(7.3 \pm 0.2)\ cm^2/Vs$ and a subthreshold slope of $(0.12 \pm 0.02) V/decade$. Microscopic images of the channel as obtained by KPFM in amplitude modulation mode are shown in Figure 1c and e.

In the height map one clearly recognizes the channel and the source and drain electrodes. The $V_{KPFM}$ map was obtained with $V_G = -2.7V$ and $V_D = 0.1V$. Accordingly, the channel shows a negative potential confirming the penetration of the gate field and absence of charge carriers. In figure 2d we report both the $V_{KPFM}$ and the height profile along the red line in figures 2c and 2e. For comparison also the KPFM base line obtained at $V_G = 0V$ is shown in the profile. The KPFM data shows that close to the contacts electrostatic interactions with the source electrodes limit the measurements, but inside the channel a flat and constant potential is measured.

Next, we continue with KPFM spectroscopic measurements done at a single position in the center of the channel to determine the IGZO's DOS parameters. Figure 3a shows the measurement of $V_{KPFM}$ obtained in the channel while the gate voltage $V_G$ was swept from -1$V$ to 5$V$. On the source electrode we obtain a constant value of 0.02V and the value is indicated in the plot to mark the flat band potential condition at $V_{G,FB} = (2.8 \pm 0.2)V$. With this condition, the model fits well and uniquely to the experimental data yielding a band tail width $E_t = (44.4 \pm 0.1)meV$ and a doping density of $N_D = (1.4 \pm 0.6) \times 10^{12} cm^{-3}$. The values are stable and variation of experimental parameters such as KPFM tip bias amplitude $V_{AC}$ or sweep time do not have a significant impact on the obtained values as shown in figures 3b and 3c. In this regard, the stability towards variations in $V_{AC}$ confirms also that the CPD interpretation of KPFM measurements in semiconductors is correct and linear response is maintained.[13]

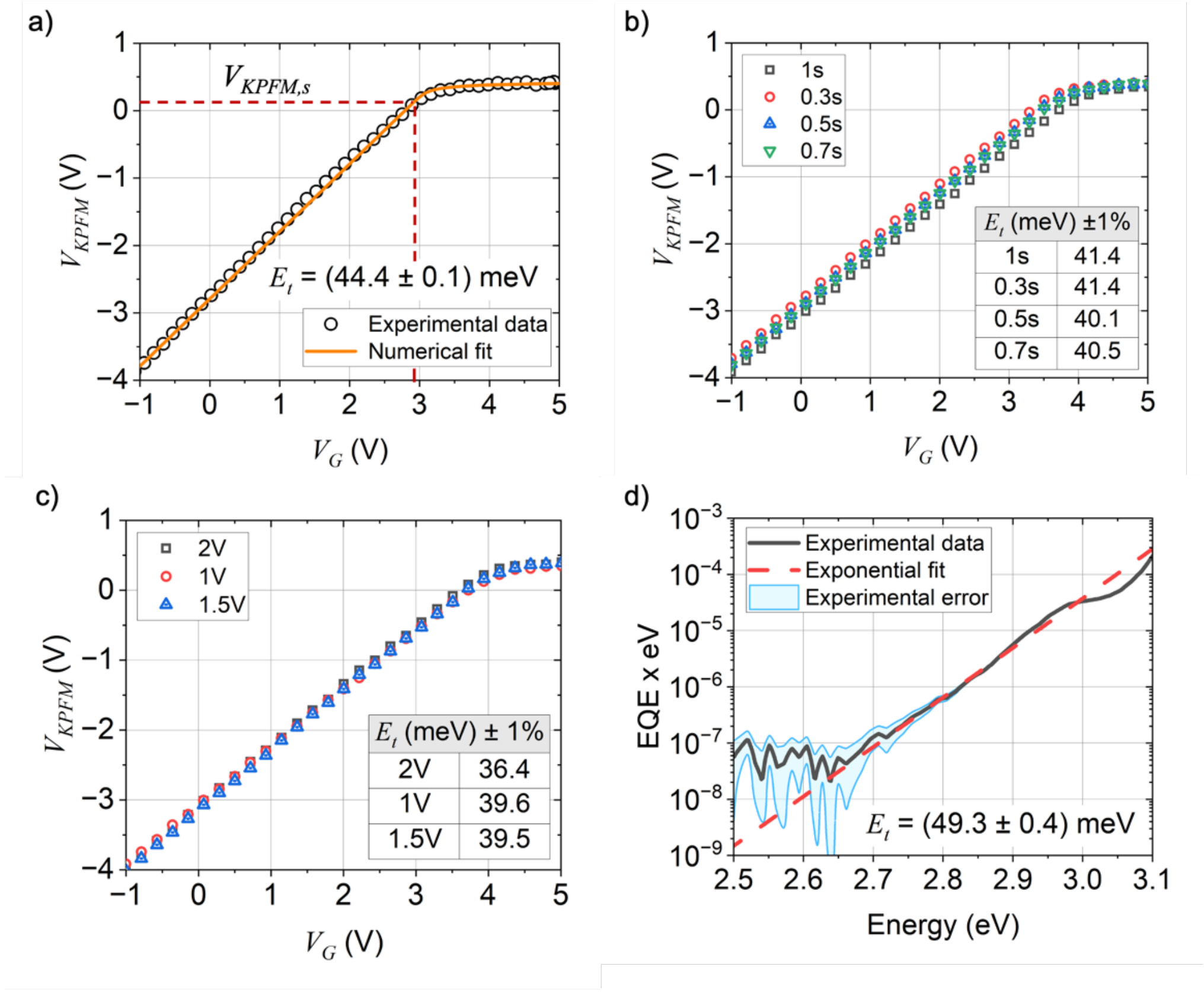

**Figure 3: KPFM spectroscopy and fit to model: (a) Example of experimental $V_{KPFM}$ vs $V_G$ curve and fit to model with extracted value of band tail width $E_t$. The red dashed line shows the KPFM potential measured on the source electrode and allows to extract the flatband potential. (b) Plot of the KPFM curve for different values of sweep time and fixed AC driving voltage with fitted values of band tail width (inset). (c) Plot of KPFM curve for different values of AC driving voltage and fixed sweep time of 1s. The fitted values of $E_t$ are reported in the inset table. (d) Plot of photocurrent external quantum efficiency (EQE) spectrum measured on the IGZO thin film transistor. The value of band tail width is comparable with the one found from KPFM.**

The extracted DOS parameters can be compared to values obtained with different, not microscopic techniques as done in table I. Our estimate of band tail width $E_t$ and doping density $N_D$ are lower than values found in literature with capacitance measurements. We note that differences in the fabrication protocol of the IGZO film might explain such variations.[23] In particular, the amount of oxygen vacancies density strongly depends on the conditions during the semiconductor deposition and annealing. For instance, a reduced number of oxygen vacancies is expected in a more crystalline IGZO sample[24]. To provide an alternative characterization of the conduction band tail in the tested transistor structures, we conduct photocurrent spectroscopy experiments. Figure 3d reports the external quantum efficiency ($EQE$) in photocurrent measured at different photonic excitation energies. A linear dependence is observed, on the logarithmic plot, and fits to an exponential with characteristic energy of $E_t(PC) = (49.3 \pm 0.4)\ meV$. This finding is in excellent agreement with the KPFM value. A slightly higher value from photocurrent spectroscopy is expected due to the contribution of the valence band tail.

| **Experiment** | **Type** | $\boldsymbol{N_D}(\mathbf{cm^{-3}})$ | $\boldsymbol{E_t}(\mathbf{meV})$ |
|---|---|---|---|
| This work | KPFM | $(1.1 \pm 0.6) \times 10^{12}$ | $44.4 \pm 0.1$ |
| Tsuji et al | CV | $1.2 \times 10^{17}$ | 80 |
| Jeon et al | Simulated | $5 \times 10^{16}$ | 67 |
| This work | Photocurrent | / | $49.3 \pm 0.4$ |

**Table I: Comparison of DOS parameters extracted from both this work and literature with different experimental techniques.** [25,26]

Finally, we demonstrate the relevance of our technique for the understanding of TFT stability and degradation. An example is the X-ray induced radiation damage causing a shift to negative voltages.[27] Such an effect is critical for different radiation related applications such as degradation of active-matrix detector backplanes or the sensitivity of microelectronic dosimeters. Despite this relevance, two basic mechanisms can potentially explain the negative shift. On the one hand, one can assume that ionizing radiation impinges on the semiconductor and causes the generation of oxygen vacancies acting as additional doping sites. Consequently, more carriers are in the channel at flat band potential and a more negative gate voltage is needed to switch-off the channel. On the other hand, ionizing radiation is known to cause positive space-charge build-up in wide band-gap oxide dielectrics. The reason is that the X-ray generated hole and electron carriers behave differently in these dielectrics. Electrons are mobile and ultimately exit from the dielectric to a conductive contact. Hole charges instead are trapped and continue to accumulate with increasing exposure dose. The generated positive space charge causes a potential offset in the gate dielectric and shifts the flat band potential to negative values.

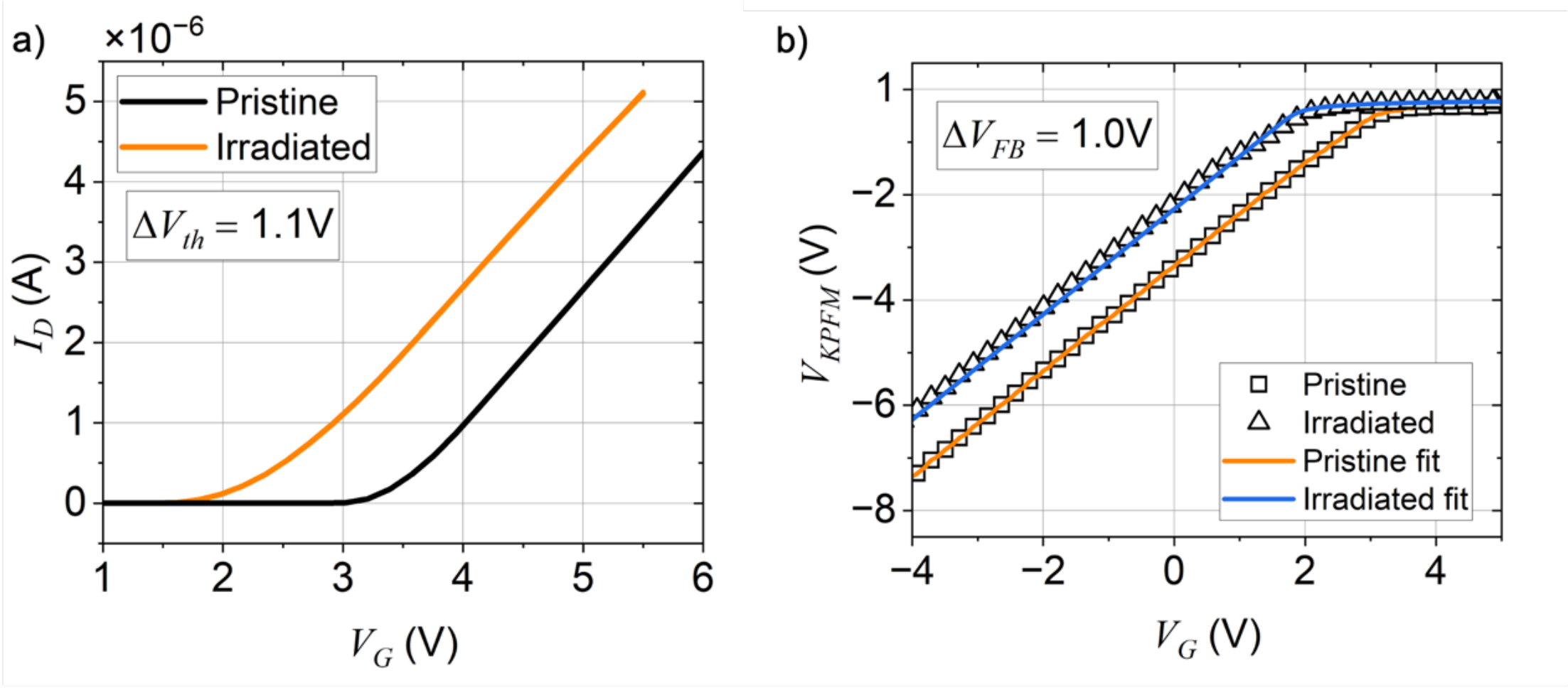

**Figure 4: Investigation of X-ray radiation damage in IGZO TFT. (a) Transfer characteristics before and after exposure to x-ray radiations (400mGy at energy correspondent to Mo-$K_\alpha$ peak, E=17.5keV). (b) KPFM spectroscopy performed in the center of the channel before and after radiation exposure. The solid lines show the fit to the developed KPFM model.**

To decipher which of the two possible causes is relevant, we perform KPFM experiments on TFTs before and after the exposure to a total dose of 400mGy. The transfer characteristics of the device, reported in figure 4a, in the pristine and irradiated condition show a clear shift in threshold of $\Delta V_{th} = (-1.10 \pm 0.01)V$. The corresponding KPFM spectroscopy measurements are shown in figure 4b together with the fits to our model. The extracted parameters are reported in table II. We note that both $N_D$ and $E_t$ are not modified during the exposition while flat band potential $V_{FB}$ is reduced. The shift is $\Delta V_{FB} = (-1.0 \pm 0.2)V$ which is coherent with the measured shift in threshold voltage. We can conclude that the DOS parameters are not modified by x-ray irradiation, therefore the effect of x-rays in our device is to generate positive charges in the gate dielectric.

| Parameter | Pristine | Irradiated |
|---|---|---|
| $N_D(\times 10^{12}\ \mathrm{cm}^{-3})$ | $1.4 \pm 0.6$ | $1.5 \pm 0.6$ |
| $E_t$ (meV) | $44.4 \pm 0.1$ | $45.8 \pm 0.2$ |
| $V_{FB}$ (V) | $2.8 \pm 0.1$ | $1.8 \pm 0.1$ |

**Table II: Comparison of density of states parameters before and after x-ray irradiation.**

**Conclusions**

Our work introduces a novel and precise method for analyzing amorphous semiconductor band tails through KPFM spectroscopy. The method relies on the solution of the non-linear Poisson equation for a metal-insulator-semiconductor (MIS) junction interacting with the AFM tip. The provided analytical solution enables the direct fit of experimental data for semiconducting films with finite thickness and the determination of DOS parameters, including band tail width ($E_t$) and doping density ($N_D$), as well as the flat band potential. We demonstrate our microscopic method on IGZO thin film transistors and obtain a local band tail width of $E_t = (44.4 \pm 0.1)meV$ that is comparable to macroscopic photocurrent measurements. We showcase the utility of our method by investigating the origin of radiation induced threshold voltage shifts in IGZO TFTs. The KPFM spectroscopy results provide clear evidence that ionizing radiation does not impact on doping, but causes static charge accumulation in the gate dielectric. The developed methodology paves the way towards microscopic imaging of DOS parameters in amorphous semiconductor thin films and provides a crucial tool to understand degradation and device physics in semiconductor thin film devices. The analytical solution allows rapid data analysis as needed in imaging experiments. The method can be combined with high resolution KPFM methods such as sideband or heterodyne techniques in order to assess the DOS also at device boundaries which are expected to be crucial for non-ideality effects.

**Materials and Methods**

We fabricated a thin film transistor (TFT) with a bottom-gate and top-contact structure, using a Corning Eagle glass substrate and patterning the desired structure by photolithography. We deposited a multilayer, multicomponent dielectric consisting of 80nm of $Ta_2O_5$ and 20nm of $Al_2O_3$ using thermal atomic layer deposition (ALD). The 60nm IGZO layer was sputtered with a composition ratio of 2.5:1.2:1 for In:Ga:Zn onto the dielectric layer. To activate the IGZO layer, we annealed the sample in air at 180 degrees Celsius. All the electrodes are made of 60nm molybdenum layer deposited by magnetron sputtering. We conducted Kelvin Probe Force Microscopy (KPFM) measurements using the Park NX-10 Atomic Force Microscope (AFM). For this experiment, we employed an NSC36 cantilever made of n-type silicon and coated with a layer of chromium and gold. The cantilever has a typical force constant of 2 N/m and resonates at a frequency of approximately 130 kHz. We acquired the transfer characteristics with a Keysight B2912A measurement unit. We kept 0.1V on the drain during every measure to maintain the linear regime, thus we obtained transistor parameters by fitting the transfer characteristics to the standard metal oxide semiconductor FET model for the corresponding regime. X-ray radiation was generated by a molybdenum tube, which we operated at a voltage of 60 kVp and filament current of 395 μA. The sample was positioned at 20cm from the shutter to achieve a doserate of 60 μGy/s delivering a total dose of 400mGy. Photocurrent spectroscopy of IGZO TFTs was done with a Xenon lamp coupled to a monochromator to select a single wavelength at a time. The TFT was biased at $V_D$=0.1V and $V_G$=0.1V. The resulting photocurrent was amplified using a Femto current amplifier (DLPCA-200) and measured with a lock-in amplifier.

# Supplemental material

## Accurate determination of band tail properties in amorphous semiconductor thin film with Kelvin Probe Force Microscopy

**Full derivation of MIS junction model with exponential DOS interacting with KPFM tip.**

Figure 1 shows a scheme of the energy levels of relevance for the MIS junction and the AFM tip. Energy levels at flat band potential are drawn in grey while we use black to depict energy levels in depletion. The diagram visualizes all parameters that are of relevance for our model. We split the derivation into two parts. First, we investigate in detail the band-bending occurring in the semiconductor. Second, we relate the band bending to the experimental observables, that is the gate voltage $V_G$ and the KPFM voltage $V_{KPFM}$. Here we derive the model for a n-type semiconductor.

**Band bending in a semiconductor with exponential DOS**

The amount of band bending in the semiconductor is quantified by the by the function $\phi(x)$, which determines the shift of the vacuum level in the semiconductor starting from the interface with air (x=0) and ending at the interface with the gate dielectric (x=t).

In the first step we derive the differential equation that determines $\phi(x)$, and its relation to the DOS of the semiconductor. Changes in the vacuum level are always attributed to the build-up of space charge $\rho$ in the semiconductor as defined by the Poisson equation:

$$\frac{d^2\phi(x)}{dx^2} = -\frac{\rho(x)}{\varepsilon\varepsilon_0} \tag{1}$$

where the specific permittivity of the semiconductor is expressed as $\varepsilon$.

The space charge results directly from the Fermi-Dirac integral

$$\rho(E_F) = N_D - q_0 \int_{-\infty}^{\infty} g(E) f(E, E_F) dE \tag{2}$$

where $g(E) = N_0 \exp\left(\frac{E}{E_t}\right)$ is the assumed density of states of the conduction band tail, $f(E, E_F)$ is the Fermi-Dirac distribution and $N_D$ the oxygen vacancies doping density, which quantifies the number of positively charged dopant sites present in the band tail. We note that here the signs in the derivation are adapted for a n-doped semiconductor with exponential tail of the conduction band.

To solve the Fermi-Dirac integral, we introduce the following simplifications: first, far from the interfaces, inside the bulk, there cannot be any space charge, thus $\rho(E_F) = 0$. Second, we set $T = 0$. We note that this assumption is less restrictive than it appears at first sight since a finite temperature values the offset can be calculated and results only in a shifting factor on the flat band voltage. We obtain

$$N_0 = \frac{N_D}{E_t} \exp\left(-\frac{E_{F,bulk}}{E_t}\right) \tag{3}$$

and then

$$\rho(E_F) = q_0 N_D \left[1 - \exp\left(\frac{E_{F,bulk} - E_F(x)}{E_t}\right)\right] \tag{4}$$

Equation (4) explains how local variations in the Fermi level are related to accumulation of space charge in the semiconductor. The variations in Fermi level are directly tied to the shift in vacuum level, indeed surface potential is defined as

$$\phi(x) = E_{F,bulk} - E_F(x) \tag{5}$$

Substituting (4) and (5) in (1) we obtain the full non-linear Poisson equation that contains DOS parameters $E_t$, $N_D$ and indirectly $N_0$:

$$\frac{d^2\phi(x)}{dx^2} = \frac{q_0 N_D}{\varepsilon\varepsilon_0}\left[1 - \exp\left(\frac{\phi}{E_t}\right)\right] \tag{6}$$

Equation (6) is a non-linear, second order differential equation which is not solvable analytically. A determined function $\phi(x)$ solves the differential equation if appropriate boundary conditions are defined. For the case of the KPFM experiment Cauchy boundary conditions are defined:

$$\begin{cases} \phi(x=0) = \phi_0 \\ \dfrac{d\phi(x=0)}{dx} = 0 \end{cases} \tag{7}$$

The first condition in (7) accounts for the finiteness of the semiconducting layer. Hence an electric field applied by the gate is not completely screened in the semiconductor, but leaves a residual causing an offset potential $\phi_0$ at the semiconductor interface exposed to the AFM tip. It is this residual potential that is related to the KPFM voltage as discussed below. The second boundary condition implies the absence of an electric field. This condition is justified by the KPFM experiment: the tip DC voltage is adjusted until no electrostatic forces are acting. In other words, the electric field is zero in the space between tip and semiconductor surface.

To solve equation (6) numerically with Runge-Kutta algorithm, we rewrite it as two coupled first order differential equations:

$$\vec{y} = \begin{pmatrix} \frac{d\phi}{dx} \\ \frac{d\mathcal{E}}{dx} \end{pmatrix} = \begin{pmatrix} \mathcal{E} \\ \frac{q_0 N_D}{\epsilon\epsilon_0}\left[1 - \exp\left(\frac{\phi}{E_t}\right)\right] \end{pmatrix} \tag{8}$$

We use a 4th order Runge-Kutta method with automatic adjustment of the step-size, to propagate the solution from initial point $\overrightarrow{y_0} = \binom{\phi_0}{0}$. Step size adjustment is crucial since the more $\phi_0$ increases the more the steepness of the solution at $x = t$ is high and the more the step-size must be reduced to correctly propagate the solution.

It is also possible to obtain analytical solutions for equation (6) for two limiting conditions: the first regards accumulation and is valid for $\phi(x = 0) > E_t$. In this condition the exponential term is predominant in the differential equation. By substituting $\alpha = \frac{q_0 N_D}{\epsilon\epsilon_0}$, equation (6) simplifies to:

$$\frac{d^2\phi}{dx^2} = -\alpha \exp\left(\frac{\phi}{E_t}\right) \tag{9}$$

The full mathematical derivation of equation (9) is reported in appendix. The final solution is then

$$\phi(x) = E_t \log\left\{\exp\left(\frac{\phi_0}{E_t}\right)\left(\tan^2\left[\frac{1}{\sqrt{2}}\exp\left(\frac{\phi_0}{2E_t}\right)\frac{x}{\lambda}\ \right] + 1\right)\right\} \tag{10}$$

Where boundary conditions (7) are already included and $\lambda = \sqrt{\frac{E_t}{q_0\alpha}}$ is a characteristic length of the semiconductor. Equation (10) describes the surface potential in function of the position inside the semiconducting layer. It can be noted that it depends also on values $\phi_0$, $N_D$ and $E_t$ through $\lambda$. As shown in Figure 2, the solution has several singularities due to the presence of the tangent function. The physically meaningful interval regards $\phi_0$values that range from $\phi_0 > E_t$ to the first singularity. The singularity occurs as a thin layer of charge accumulation appears at the gate-dielectric/semiconductor interface, that screens any further increase in gate field.

The second limiting analytical solution for the non-linear Poisson's equation (6) is obtained in depletion condition ($\phi_0 << E_t$):

$$\phi(x) = -\frac{\alpha}{2}x^2 + \phi_0 \tag{11}$$

The predicted relation between $\phi_t = \phi(x = t)$ and $\phi_0$ is linear once the thickness of the layers is fixed, as expected for depletion regime.

**Relation of band bending to experimental parameters**

In KPFM experiments we measure the relation between the gate voltage $V_G$ and the variation in KPFM tip voltage, namely $V_{KPFM}$. The latter, $V_{KPFM}$, is directly related to $\phi_0$ through the relation $\phi_0 = q_0 V_{KPFM} - \phi_{ds}$, as it can be deducted directly from Figure 1. $\phi_{ds}$ denotes additional offsets that originate from (i) possible dipole layer forming at the semiconductor interface exposed to the AFM tip (ii) workfunction of the AFM tip and (iii) amplifier offsets. The direct relation between $\phi_0$ and $V_{KPFM}$ is supported by the CPD interpretation of KPFM experiments. The measured potential in a KPFM experiment can be seen as an approximation of the difference between the work functions of tip and sample. In semiconducting materials, the presence of an external electric field influences the internal charge distribution. It induces the formation of a region with non-zero net charge, the so-called space charge density. In general, the CPD interpretation of KPFM measures is valid for materials with a fixed charge distribution, thus with a charge per unit area that can be expressed as $\sigma = C(V - V_{CPD})$.

In order to relate the gate voltage $V_G$ to band bending, we consider again the energy level diagram in Fig.1 and write down two equations: one for flat band condition and one for depletion:

$$\begin{cases} q_0 V_{FB} = \Phi_s - \Phi_g + \phi_{ds} + \phi_{dg} \\ q_0 V_G = \Phi_s - \Phi_g + \phi_{ds} + \phi_t + q_0 V_{ox} + \phi_{dg} \end{cases}$$

Here $\phi_t$ denotes the potential at the semiconductor/dielectric interface $\phi_t = \phi(x = t)$. If we subtract the second equation from the first and divide for the electron charge, we obtain an equation that relates the gate voltage to $\phi_0$:

$$V_G = V_{FB} + \frac{\sigma(\phi_0)}{c_{ox}} + \frac{\phi_t(\phi_0)}{q_0} \tag{12}$$

Here $\sigma$ is the value of accumulated charge in the semiconductor and determines the potential drop across the dielectric $V_{ox}$ determined by $c_{ox}$ , the capacitance per unit area of the dielectric. $V_{FB}$ is the flat band potential and contains different physical contributions such as the gate metal work function or interfacial dipole layers causing a shift in vacuum level. Moreover, flat band potential does depend on other DOS parameters, namely $N_D, E_t, N_0$ and the experimental offsets, as it is explained below.

To obtain the full relation to the experimental parameters, we have to express $\sigma$ as a function of $\phi_0$ using the calculated band bending, with either numerical or approximated analytical scheme. Starting directly from Poisson's equation (6) we can write

$$\frac{d^2\phi}{dx^2} = \alpha\left[1 - \exp\left(\frac{\phi}{E_t}\right)\right]$$

with $\alpha = \frac{q_0 N_D}{\epsilon\epsilon_0}$. Multiply both terms for $\frac{d\phi}{dx}$ and integrate. The result is

$$\left(\frac{d\phi}{dx}\right)^2 = 2\alpha\left[\phi - E_t \exp\left(\frac{\phi}{E_t}\right)\right] + c$$

which is a solution for the first derivative of the surface potential. We can again impose boundary conditions (7) to find the correct value of constant c.

$$\frac{d\phi}{dx} = \sqrt{2\alpha\left\{\phi - \phi_0 - E_t\left[\exp\left(\frac{\phi}{E_t}\right) - \exp\left(\frac{\phi_0}{E_t}\right)\right]\right\}} \tag{13}$$

From Poisson's equation it is also known that the amount of accumulated charge in the semiconductor can be expressed in function of the first derivative, thus

$$\sigma = \epsilon\epsilon_0\left[\left(\frac{d\phi}{dx}\right)_{x=t} - \left(\frac{d\phi}{dx}\right)_{x=0}\right] = \epsilon\epsilon_0\sqrt{2\alpha\left\{\phi_t - \phi_0 - E_t\left[\exp\left(\frac{\phi_t}{E_t}\right) - \exp\left(\frac{\phi_0}{E_t}\right)\right]\right\}} \tag{14}$$

The last required parameter to completely explicit equation (12) if $V_{FB}$. Flat band potential is not a material-related parameter, but it is important when dealing with semiconductor-based devices since it indicates, in transistors, the switching point between depletion and accumulation condition. Consider that $q_0 V_{FB} + \Phi_g + q_0\phi_{dg} = \phi_{ds} + \Phi_s$ and at flat potential it is also true that $\phi_{ds} + \Phi_s = E_{F,bulk}$. Recalling equation (3) it is possible to express $E_{F,bulk}$ as

$$E_{F,bulk} = -E_t \log\left(\frac{N_D}{N_0 E_t}\right)$$

Combining all the previous equations we obtain

$$q_0 V_{FB} = \Phi_g + \phi_{dg} - E_t \log\left(\frac{N_D}{N_0 E_t}\right) \quad (15)$$

Relation (15) shows, as anticipated upwards, that flat band potential is not a proper DOS parameter but depends on doping densities, band tail slope, metal work function and potential offsets of the system. Combining equations (10), (12) ,(14) and (15) we obtain an expression that relates gate voltage and $\phi_0$, that is the KPFM potential. The relation is then used to fit the experimental data. For the approximate analytical model we obtain:

$$V_G = \phi_t + \frac{\epsilon\epsilon_0}{C_{ox}}\sqrt{\frac{2\alpha}{q_0}\left\{\phi_t - \phi_0 - E_t\left[\exp\left(\frac{\phi_t}{E_t}\right) - \exp\left(\frac{\phi_0}{E_t}\right)\right]\right\}} + \frac{\Phi_g}{q_0} + \frac{\phi_{dg}}{q_0} - \frac{E_t}{q_0}\log\left(\frac{N_D}{N_0 E_t}\right) \quad (16)$$

Figure 3 compares results calculated with the numerical and the approximate analytical scheme. One can note that the value of $\sigma$ found with the analytical relation corresponds to the one found with the numerical approach. Only when the semiconductor approaches the flat band situation, small differences become notable as the condition $\phi(x = 0) \gg E_t$ is no longer valid and the analytical solution starts to fail. However, this discrepancy is negligible when dealing with the full experimental $V_{KPFM}$ vs $V_G$ curve, as it can be seen in Figure 3b. Its only effect is to produce an incorrect slope in depletion region. Therefore, as expected, the approximated solution could not be reliable in depletion condition, where the correct behavior is provided by solution (11).

Looking at the band diagram in Figure 1 one can detect the existence of two offsets: $\phi_{dg}$ and $\phi_{ds}$ which enters directly in equation (16). These are strictly related to the experimental measure, and it is necessary to clarify their relation with the numerical solution. The first offset $\phi_{dg}$ is related to the formation of a dipole between the gate and the dielectric whose presence produces a shift in gate voltage values. The second offset $\phi_{ds}$ can be due either to the formation of a dipole at the air – semiconductor interface or the non – perfect alignment of source and semiconductor. It produces an unknown shift in KPFM potential. The interaction with air can be deleted by performing the measure in Argon atmosphere, while the second one cannot be cancelled since it's dependent on the fabrication process of the device.

In order to fit equation (16) to experimental results one has to consider the different fitting parameters: $E_t$, $N_D$, $\phi_{ds}$, $\phi_{dg}$ and their interdependency. Oxide capacitance $C_{ox}$, $\epsilon$ and thickness t are typically known for a measured structure. The presence of interdependencies between parameters is the cause of the indeterminacy of the system, indeed it is not possible to find a unique set of DOS parameters only by fitting the experimental $V_{KPFM}$ vs $V_G$ curve as many combinations are possible. Because of that, the determination of the two unknown offsets cited upwards is important as they provide an uncertainty directly on the estimation of some parameters, such as trap doping density $N_D$. To solve this issue, it is possible to distinguish two cases: if flat band voltage is known, then it is enough to adjust the KPFM curve knowing that $q_0 \Delta V_{KPFM}(V_G = V_{FB}) = \phi_{ds}$. Second, if one can assume the the dipole $\phi_{ds}$ is zero, the flat band potential is the voltage that satisfies the condition $q_0 \Delta V_{KPFM}(V_G = V_{FB}) = 0$. We can actually assume $\phi_{ds} = 0$ for our system since the potential barrier between source and semiconductor is expected to be low. If we suppose to have an error of $2k_B T \simeq 50meV$ on the measure, due to the uncertainty on $\phi_{ds}$, it is possible to propagate it to the fitting parameters. We obtained an error of 0.6% on the estimate of $E_t$ and a wider range for $N_D$. Specifically, for $N_D = 2.01 \times 10^{15}$ we obtained a maximum possible value $N_D^{max} = 4.02 \times 10^{15} cm^{-3}$ and a minimum $N_D^{min} = 1.04 \times 10^{15} cm^{-3}$. As expected, the value of $E_t$ is not influenced by $\phi_{ds}$, which plays a role only in the correct determination of trap states doping density.

**Appendix: full mathematical derivation of approximated solutions**

Equation (6) where can be solved analytically for the case of strong accumulation ($|\phi(x = 0)| \gg E_t$) by multiplying both members for $\frac{d\phi}{dx}$.

$$\frac{d\phi}{dx}\frac{d^2\phi}{dx^2} = \alpha \exp\left(\frac{\phi}{E_t}\right)\frac{d\phi}{dx}$$

Integrate now both terms. The first one becomes

$$\int \frac{d\phi}{dx}\frac{d^2\phi}{dx^2} dx = \left(\frac{d\phi}{dx}\right)^2 - \int \frac{d\phi}{dx}\frac{d^2\phi}{dx^2} dx \Longrightarrow$$
$$\int \frac{d\phi}{dx}\frac{d^2\phi}{dx^2} dx = \frac{1}{2}\left(\frac{d\phi}{dx}\right)^2$$

The integration of the right member leads to

$$\int \alpha \exp\left(\frac{\phi}{E_t}\right)\frac{d\phi}{dx} dx = \alpha E_t \exp\left(\frac{\phi}{E_t}\right) + c$$

By imposing the second boundary conditions in (7) one can obtain

$$0 = 2\alpha E_t \exp\left(\frac{\phi_0}{E_t}\right) + c$$

$$\frac{d\phi}{dx} = \pm\sqrt{2\alpha E_t \left[\exp\left(\frac{\phi}{E_t}\right) - \exp\left(\frac{\phi_0}{E_t}\right)\right]} \qquad (A1)$$

An analytical solution to (A1) can be found and results to be

$$\phi(x) = E_t \log\left\{\exp\left(\frac{\phi_0}{E_t}\right)\left(\tan^2\left[\frac{1}{2}\exp\left(\frac{\phi_0}{2E_t}\right)\left(c_1 + \sqrt{2}\,\frac{x}{\lambda}\right)\right] + 1\right)\right\} \qquad (A2)$$

where $\lambda^2 = \frac{E_t}{\alpha}$. By using again the boundary conditions (7) one can obtain the full approximated solution to the Poisson's equation

$$\phi(x) = E_t \log\left\{\exp\left(\frac{\phi_0}{E_t}\right)\left(\tan^2\left[\frac{1}{2}\exp\left(\frac{\phi_0}{2E_t}\right)\sqrt{2}\frac{x}{\lambda}\right] + 1\right)\right\} \qquad (A3)$$

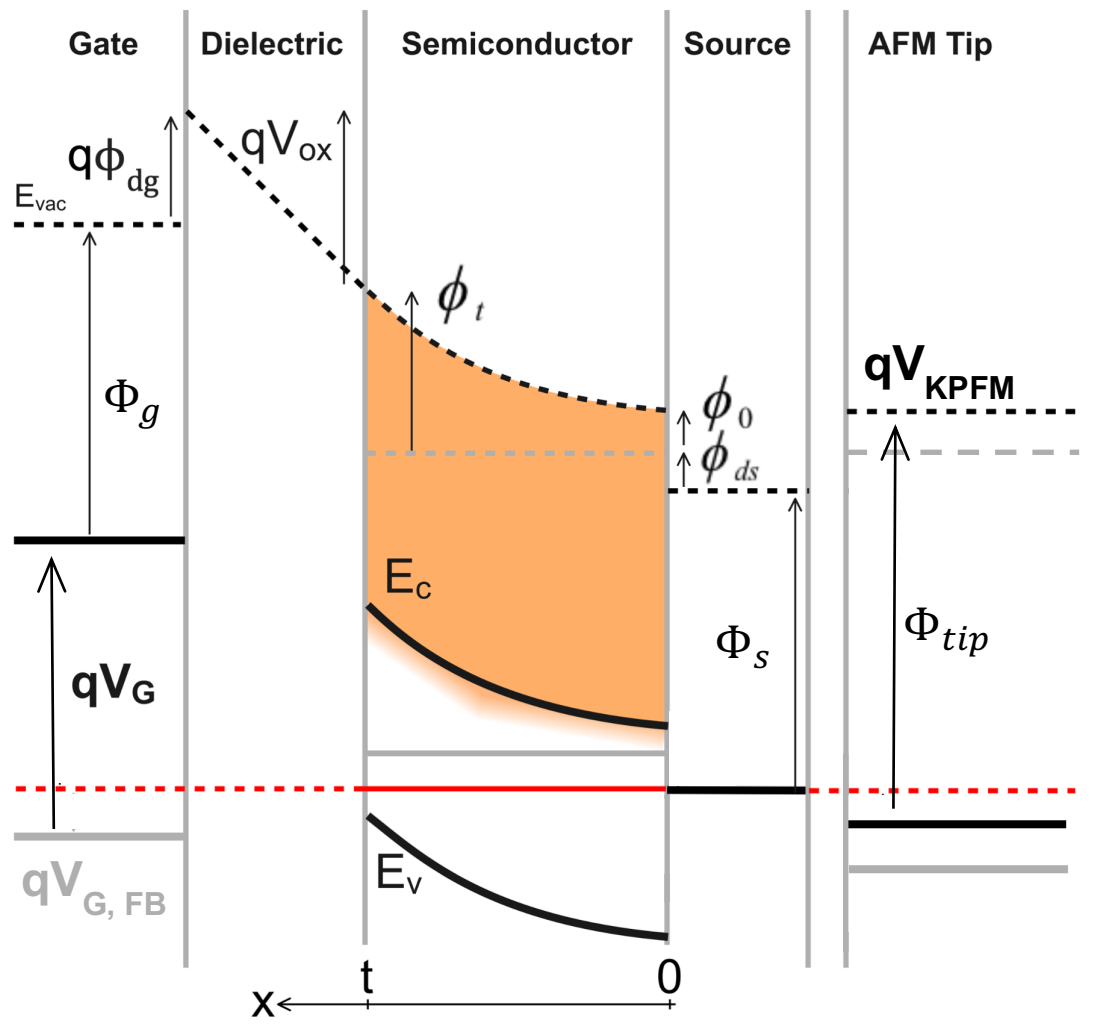

Figure 1: Schematic band diagram of a MIS junction. Depletion condition is drawn in black while flat band condition is depicted in grey. The red horizontal line represents Fermi level of the semiconductor, which is grounded, thus fixed.

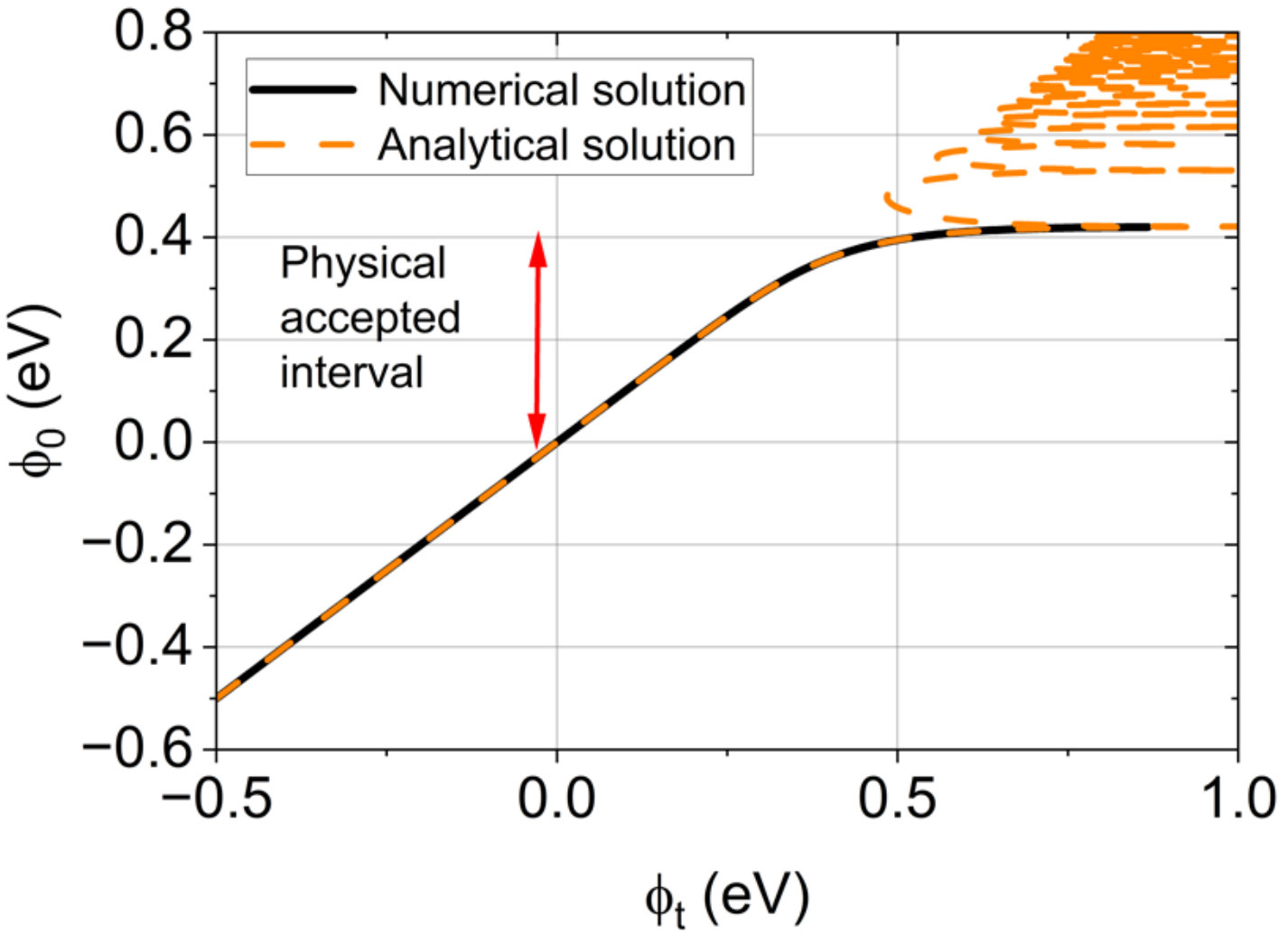

Figure 2: Approximated analytical solution for accumulation regime and comparison with numerical solution. Parameters of the simulation are $N_D = 10^{13} cm^{-3}, E_t = 50 meV, \ t = 60 nm$ .With the red arrow it is depicted the range of validity of the analytical solution: it is valid only for $\phi_0 > E_t$ and up to the first tangent singularity.

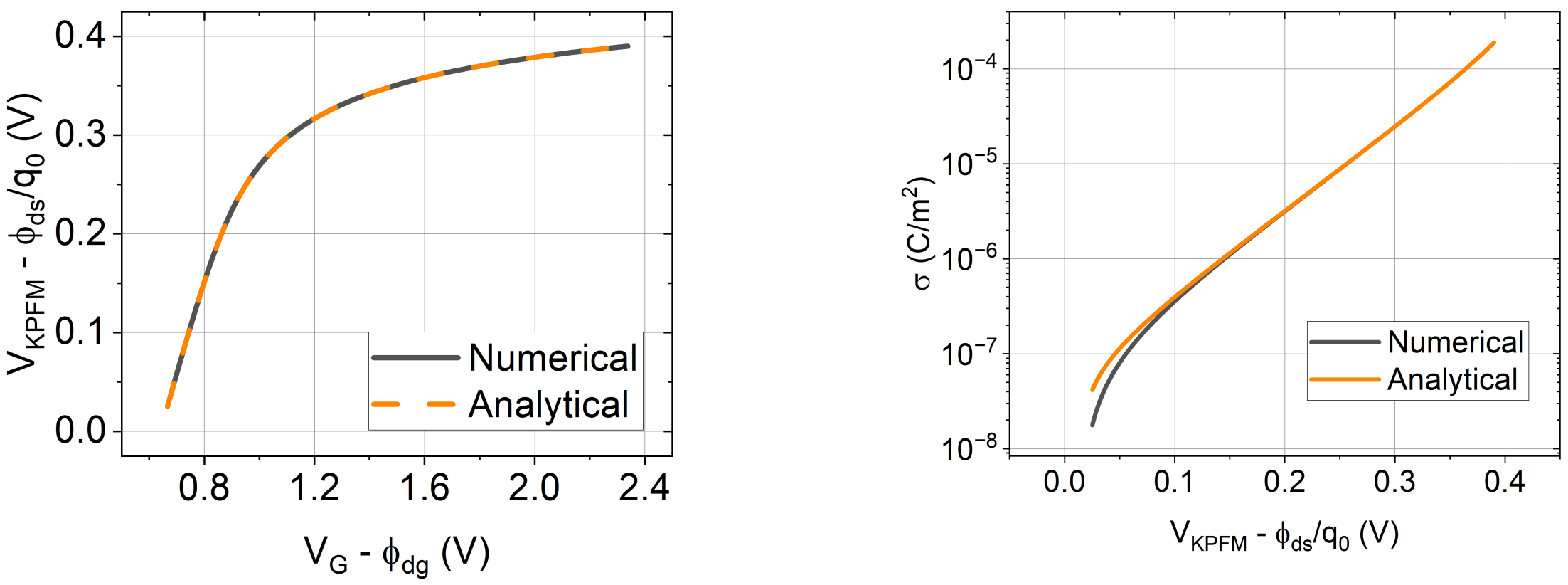

Figure 3: Plot of accumulated charge density in the semiconductor calculated with both numerical and analytical model. (right) Plot of $\phi_0$ in function of gate voltage. (left) Both simulations were done only in accumulation regime with the following parameters: $N_D = 10^{14} cm^{-3}$ , $E_t = 60 meV$ , $t = 50 nm$.